# A Framework for Providing E-Services to the Rural Areas using Wireless Ad Hoc and Sensor Networks

Al-Sakib Khan Pathan[1], Humayun Kadir Islam[2], Sabit Anjum Sayeed[3], Farruk Ahmed[2], Choong Seon Hong[1]
[1] {spathan@networking.khu.ac.kr, cshong@khu.ac.kr}, Department of Computer Engineering, Kyung Hee University, Korea
[2] {humayun, farruk}@northsouth.edu, Department of Computer Science & Engineering, North South University, Bangladesh
[3] sabit_a_s@yahoo.com , University of Ottawa, Canada

*Abstract*-In recent years, the proliferation of mobile computing devices has driven a revolutionary change in the computing world. The nature of ubiquitous devices makes wireless networks the easiest solution for their interconnection. This has led to the rapid growth of several wireless systems like wireless ad hoc networks, wireless sensor networks etc. In this paper we have proposed a framework for rural development by providing various e-services to the rural areas with the help of wireless ad hoc and sensor networks. We have discussed how timely and accurate information could be collected from the rural areas using wireless technologies. In addition to this, we have also mentioned the technical and operational challenges that could hinder the implementation of such a framework in the rural areas in the developing countries.

## I. INTRODUCTION

Mobile computing, grid computing, pervasive micro-sensing and actuation [1], and recent advancements in wireless technologies might change our surrounding environment in a way that we haven't yet imagined. In addition to the development of rich and developed areas, wireless systems could significantly contribute to the development of technologically lagging rural areas especially in the developing countries. In recent years, various wireless technologies have shown promise for various futuristic public applications. With the proliferation of mobile computing devices like laptops, palmtops, personal digital assistants (PDAs), and plummeting costs of telecommunication devices; various wireless systems and concepts like, wireless ad hoc networks, wireless sensor networks, ubiquitous computing, grid computing etc. have been introduced. These emerging technologies could effectively be used for smartening the environment as well as for improving the socio-economic status of the rural areas in the developing countries.

In this paper we present a detailed framework for providing various e-services like, e-learning, e-academics, e-medicine and e-health care, e-business etc. to the rural areas using wireless sensor and wireless ad hoc networks, which could also be helpful for connecting the hard-to-reach rural areas and the government more efficiently. As timely and accurate data is essential for critical decision and policy making (especially in emergency situations), our framework ensures quick and accurate data acquisition using wireless technologies, which could definitely assist the government to take necessary strategic decisions and immediate actions. We also address the practical/operational challenges that could hinder the implementation of such a framework in the rural areas of the developing countries.

The rest of the paper is organized as follows: section II gives an overview as well as states the major benefits of wireless sensor and ad hoc networks, section III discusses various parts of our proposed framework, section IV mentions the e-services that could be provided by using the framework, section V mentions the technical/operational challenges for implementing such a framework and section VI concludes the paper.

## II. WIRELESS SENSOR AND AD HOC NETWORKS – A BACKGROUND

### A. Overview of Wireless Sensor and Ad Hoc Networks

A wireless sensor network (WSN) is a combination of a number of sensor nodes connected via wireless communications. Recent advancements in the commercial IC (Integrated Circuit) fabrication technology and wireless technologies have made it possible to integrate sensing, signal processing and wireless communication in one integrated circuit [2]. These devices are popularly known as wireless integrated network sensors (WINS) [3]. Sensors can monitor temperature, pressure, humidity, soil makeup, vehicular movement, noise levels, lighting conditions, the presence or absence of certain kinds of objects or substances, mechanical stress levels on attached objects, and other properties [4]. Their mechanism may be seismic, magnetic, thermal, visual, infrared, acoustic or radar [5].When networked, such sensor nodes could build up the part of larger systems, providing data, as well as performing and controlling multitude of tasks and functions (for example, surveillance, target tracking etc). In practical, large number of sensor nodes could be dispersed on demand at any time at designated locations, referred to as area of interest (AOI), or at random at specified areas. Fig. 1(a) shows a graphical view of a wireless sensor network.

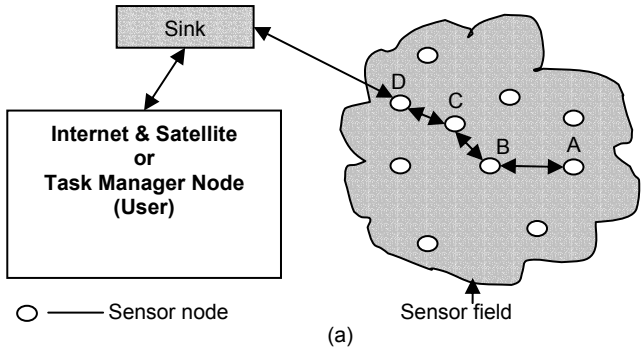 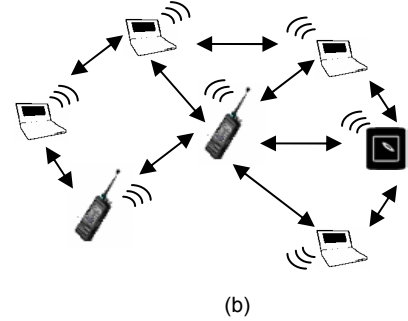

Fig. 1. (a) Sensors in a sensor network
(b) Wireless ad hoc network

Wireless ad hoc networks, on the other hand are self-organizing, dynamic topology networks formed by a collection of mobile nodes through radio links [6]. Minimal configuration, absence of infrastructure and quick deployment make them convenient for emergency situations such as natural or human-induced disasters, medical emergency or military conflicts. Some of the characteristics of ad hoc networks are: wireless connection among the nodes within the network, mobile nodes within the network, infrastructure less or semi-infrastructure, dynamically changing topology, no centralized access point, energy constrained nodes and multi-hop communication. Fig. 1(b) shows the conceptual view of a wireless ad hoc network.

*B. Benefits of Wireless Sensor Networks*

Fundamental objectives of sensor networks are reliability, accuracy, flexibility, cost effectiveness and ease of deployment. The benefits of WSNs are outlined below:
- *Sensing accuracy:* The utilization of a larger number and variety of sensor nodes provides potential for greater accuracy in the information gathered as compared to that obtained from a single sensor.
- *Area coverage:* This implies that fast and efficient sensor network could span a greater geographical area without adverse impact on the overall network cost.
- *Fault tolerance:* Device redundancy and consequently information redundancy can be utilized to ensure a level of fault tolerance in individual sensors.
- *Connectivity:* Multiple sensor networks may be connected through sink nodes (see Fig. 1(a)), along with existing wired networks (e.g. Internet). The clustering of networks enables each individual network to focus on specific areas or events and share only relevant information.
- *Minimal human interaction:* Having minimum human interaction makes the possibility of having less interruption of the system.
- *Operability in harsh environments:* Sensor nodes, consisting of robust sensor design, integrated with high levels of fault tolerance can be deployed in harsh environments that make the sensor networks more effective.
- *Dynamic sensor scheduling:* Implying some scheduling scheme, sensor network is capable of setting priority for data transmission.

*C. Benefits of Wireless Ad Hoc Networks*

Like wireless sensor networks, wireless ad hoc networks also have some attractive benefits. The significant benefits of ad hoc networks are:
- *Ease of Deployment:* Ad hoc networks are easily deployable as they do not need any fixed infrastructure of central administration.
- *Speed of Deployment:* Ad hoc networks are deployable on the fly. They are autonomous and infrastructure-less or semi-infrastructure.
- *Cost of Deployment:* There is no incremental cost for deployment; however, costs may rise depending upon the nodes associated with the network.
- *Anywhere, anytime:* Wireless ad hoc networks could be deployed anywhere, anytime especially in the hostile or geographically harsh areas where fixed network deployment is difficult.

III. PROPOSED FRAMEWORK

Considering the benefits and features of wireless sensor and wireless ad hoc networks, we propose an efficient and cost-effective framework for providing various e-services to the rural areas especially in the developing countries. Timely and accurate data is necessary for good governance. In most of the developing countries, often the governments cannot get the actual information about many of the rural areas. On the other hand, as most of the rural areas are either hard-to-reach or technologically lagging, they cannot get most of the benefits offered by the government. In some cases, some rural areas are deprived of good education facilities, health care, food and nutrition, disaster relief etc. So, acquiring exact and timely data from these rural areas is a crucial task which could assist the government to extend their development activities as well as to provide the rural areas with the facilities to fulfill the basic needs for living. In our framework, wireless technologies are used to bridge this gap between rural areas and the government. We also discuss how the collected data could effectively be used for providing various e-services to the rural areas.

*A. Framework – Phase One*

In our proposed system, wireless sensor networks are used for collecting various types of data in the phase one. In this case, sensors are deployed in the crucial parts of the rural areas like, river banks, geographically challenging parts (for example; hilly areas) and other areas of interest. The sensor networks could collect various critical data (e.g., level of water in the rivers which could help for flood warning, earthquakes etc.) and send them to the village kiosks. Each of the villages or rural areas has at least a kiosk (acts as a sink) which is equipped with computers for storing acquired data. In addition to the data collected by the wireless sensor networks, other necessary data like demographic data, health care information (for example, arsenic problem in the rural areas of Bangladesh), agricultural information, educational information etc. could be manually entered into the kiosks. Various kiosks, $K_1, K_2, K_3, \ldots, K_n$ established in different rural areas are shown in the Fig. 3. All of these kiosks are capable of wireless communications.

*B. Framework – Phase Two*

In phase two, Mobile Access Points (MAPs) play the major role. A MAP is a vehicle mounted wireless access point which uses low-cost Wi-Fi (Wireless Fidelity) technology [7], [8]. In Fig. 3. we have shown $MAP_i$ where $i$=1,2,3,4 ……n and $i \geq n$ ($n$ is the number of kiosks). These MAPs move around the kiosks in the rural areas and collect data from the kiosks. When a MAP comes near to a kiosk, a wireless ad hoc network is automatically formed and all the raw data are downloaded into the MAP. The 802.11b Wi-Fi technology operates in the 2.4 GHz range offering data speeds up to 11 megabits per second [9]. There are two other specifications that offer up to five times the raw data rate, or 54 Mbps. One is 802.11g which operates on the same 2.4 GHz frequency band as 802.11b. The other alternative 802.11a, occupies frequencies in the 5 GHz band. If offers less range of coverage than either 802.11b and 802.11g but offers up to 12 non-overlapping channels, compared to three for 802.11b or 802.11g, so it can handle more traffic than it's 2.4 GHz counterpart [10]. Based on the need or particular situation, any of these specifications is chosen for a particular area for transmitting stored data from the kiosks to the MAPs. These data are then taken and delivered to the DPCs (Data Processing Center) located at nearby towns by the MAPs. So, a Wi-Fi enabled MAP operates in two ways:

1) Forms wireless ad hoc network when comes close to the kiosks and collects data from the rural kiosks using Wi-Fi radio transceivers.

2) Again, forms wireless ad hoc network when comes close to the DPCs and delivers raw data to the DPCs using Wi-Fi radio transceivers.

The major task of the MAPs is to ensure quick acquisition and delivery of data about the rural areas and to bridge the gap between the rural areas and towns.

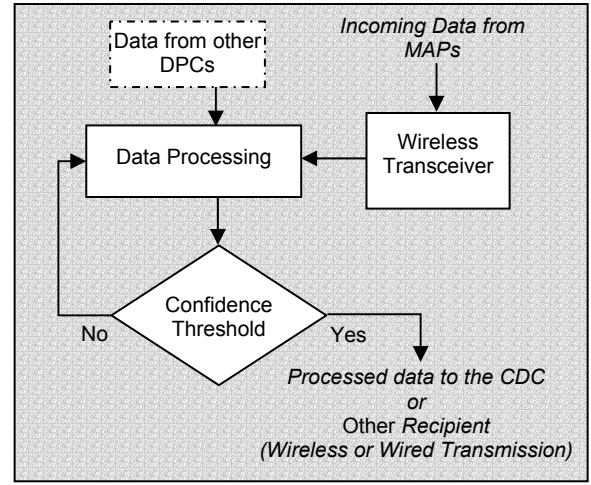

Fig. 2. Functional diagram of a DPC aided with wireless communications

*C. Framework – Phase Three*

DPCs are capable of wireless communications. All the incoming data from the MAPs are stored temporarily and processed in the DPCs. For data integrity and authenticity, all the DPCs are networked either using wireless or wired connections. In Fig. 3, we have shown the $DPC_1, DPC_2, DPC_3,...DPC_m$ where, $m \leq n \leq i$. those are set up throughout the town areas. A functional diagram of a DPC is shown in Fig. 2. In the data processing phase, related data from other DPCs are collected using wired or wireless transmission and after processing the data, confidence threshold is checked. Error detection is defined by the predetermined threshold and if necessary sent back to processing mechanism. Once processed data is ready, they are transmitted to the CDC (Central Data Center) for the next phase. Wireless or wired transmission could be used for this data transfer.

*D. Framework - Phase Four*

In this phase, the processed data from the DPCs are gathered in the CDC. Now, the CDC could combine this data with the past records for a particular rural area and supply the data to the DCC (Decision and Command Center). In this way, the government gets the timely and processed data from the rural areas and could take immediate actions accordingly. This data not only helps the government to provide various services to the rural areas but also helps in the emergency situations. Depending upon the gravity of the data, (for example, emergency situations like, cyclones, flash floods etc.), the MAPs or the DPCs could use the wireless communications to directly call the emergency or other services bypassing the CDC and DCC. Fig. 3. shows the optional wireless connectivity using dashed lines.

IV. PROVIDING E-SERVICES TO THE RURAL AREAS

Our proposed framework could effectively be used for gathering information which would definitely facilitate the delivery of several e-services to the hard-to-reach rural areas.

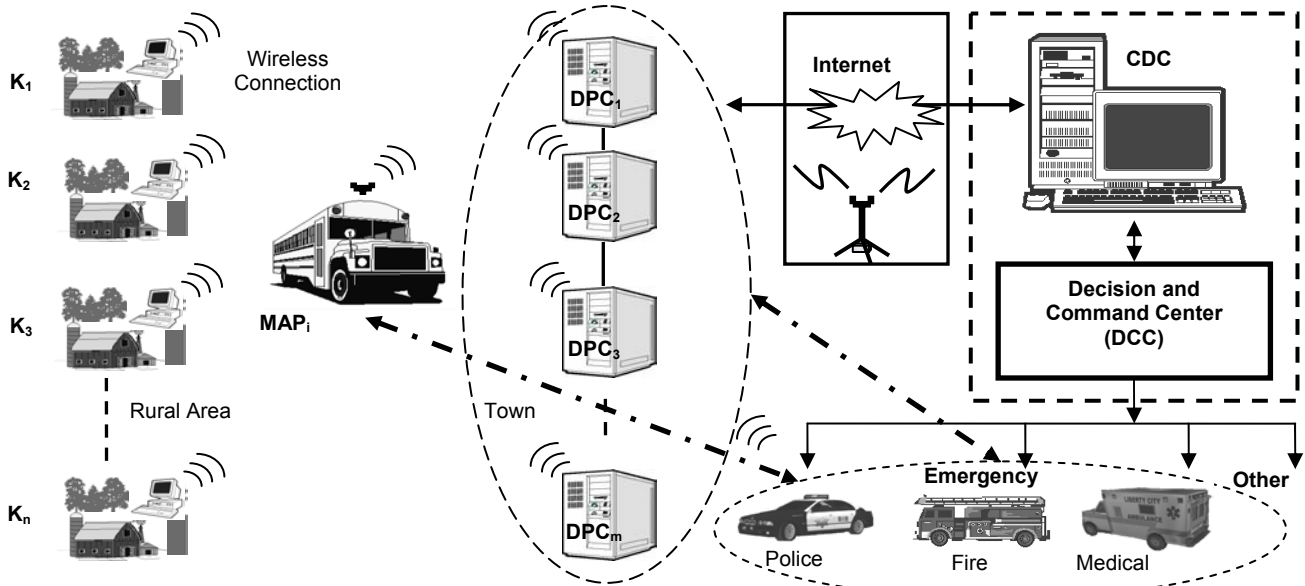

Fig. 3. Proposed Framework aided with wireless technologies

Various e-services that could be provided by the government using this framework are:

*A. E-Medicine and E-Health care*

Computer-supported medical diagnosis or e-medicine [11] and e-health care [12], [13] could significantly be benefited by the use of our framework. For improving the health scenario of the rural areas, at least a paramedical doctor should be associated with each of the kiosks or in the village dispensaries. The doctors could use the computers for keeping records about the symptoms of the diseases of the incoming patients. The MAPs could download these medical solution requests from the kiosks and deliver those to the town hospitals. The hospitals could have DPCs or nearby DPC could supply the requests to the hospital. In the reverse way, the medical solutions from the hospitals could be sent to the village dispensaries using wireless technologies. Again the government could be well-informed about some critical health information like arsenic problem, diarrhoea breakout etc. in the rural areas. This framework certainly reduces the cost of sending medical teams. Based on the information received, in severe situations, the government could take this decision but otherwise, the solutions related to these health issues could be sent to the village paramedical doctors who will in turn work accordingly. Thus, it ensures low cost e-medicine and e-health care service for the rural areas.

*B. E-Learning*

E-Learning refers to the utilization of information systems and information technology in educational services [14]. Various applications and processes that could be delivered in synchronous or asynchronous format, like web-based learning, computer-based learning, virtual classrooms, digital collaboration etc. are the examples of E-Learning methodologies. The ultimate goal of E-Learning is to bring the learning to the learners, not to bring the learners to learning.

Many rural areas do not have proper educational facilities; hence, using of the proposed framework could be used for delivering quality learning materials to the students as well as to bring quality learning to the learners. This could definitely facilitate the improvement of the education sector in the developing countries. As mentioned earlier, depending upon the requirement particular Wi-Fi technology could be chosen and the MAPs could be used for delivering e-learning materials even in audio-visual format.

*C. E-Commerce and E-Business*

Some of the rural areas have small cottage industries which produce traditional handi-crafts which often have a great demand in the other areas or sometimes in other countries. In our framework, the MAPs could be used for delivering the demands from outside and also the information about exporting these traditional goods. At the same time, information about other goods and products could be supplied to the rural areas. So, it could be a means of expanding e-commerce even in the rural areas.

V. OPERATIONAL AND TECHNICAL CHALLENGES

While our proposed framework seems to be promising for the development of rural areas, there could arise some operational and technical challenges for implementing such a framework:

- Primary installation costs might be a bit high for the developing countries to bear.
- Detailed planning is required to decide which areas should have kiosks, which areas should be chosen for deploying wireless sensors etc.
- Computer literate people are required at each village kiosk. So, some of the people should be trained for this.
- Some rural areas don't have good communication facilities. Roads are often not suitable for the movement

- of heavy vehicles like the wireless access point carrier buses or MAPs.
- In disastrous situations like, storm, heavy rainfall etc. the wireless technology might not come as useful.

## VI. DISCUSSION AND CONCLUSION

Any new system is often expensive and has some preliminary installation costs. But, once it is set up, it could run smoothly and serve for the greater benefits. As wireless technologies are growing rapidly to replace wired systems in many sectors and to make life easier, we believe that, our proposed framework could play a key role for the rural development by providing various e-services to the village inhabitants. In comparison with other existing systems, we believe that, this framework would be helpful to provide cost-effective tele-medicine facilities to the people in the rural areas. It could also be very useful to face pre and post-disaster situations.